\documentclass[12pt]{iopart}

\usepackage{iopams}
\usepackage{graphicx}
\usepackage{bm,epsfig}

\begin{document}

\title{Group velocity control in the ultraviolet domain via interacting dark-state resonances}

\author{Mohammad Mahmoudi$^{1,2}$, Mostafa Sahrai$^3$, and J\"org Evers$^1$}

\address{$^1$Max-Planck-Institut f\"ur Kernphysik, Saupfercheckweg 1, 
D-69117 Heidelberg, Germany}

\address{$^2$Physics Department, Zanjan University, P. O. Box 45195-313,
Zanjan, Iran}

\address{$^3$Research Institute for Applied Physics and
Astronomy, University of Tabriz, Tabriz, Iran}

\date{\today}

\begin{abstract}
The propagation of a weak probe field in a laser-driven four-level 
atomic system is investigated. We choose mercury as our model
system, where the probe transition is in the ultraviolet region.
A high-resolution peak appears in the optical spectra due 
to the presence
of interacting dark resonances. We show that this narrow peak
leads to superluminal light propagation with strong absorption,
and thus by itself is only of limited interest. But if in addition a weak 
incoherent pump field is applied to the probe transition, 
then the peak structure can be changed such that both sub- and superluminal 
light propagation or a negative group velocity can be achieved without
absorption, controlled by the incoherent pumping strength.
\end{abstract}


\maketitle

\section{Introduction}
Optical properties of an atomic medium can be substantially modified
by the application of external fields. In particular, atomic coherence
induced by laser fields plays an important 
role in light-matter interaction and has found numerous implementations in
optical physics~\cite{ficekbook}. One prominent application is the modification
of the propagation of a light pulse through an atomic medium,
which
depends on the dispersive properties of the medium. The study of
such pulse propagation phenomena has been triggered by a
series of papers by Sommerfeld and Brillouin~\cite{Sommer, Brill}
and continues to be of much 
interest~\cite{Harris,Schmit,Field,Xiao,agarwaldey,patnaik}. 
It is well known that the
group velocity of a light pulse can be slowed down~\cite{Hau,Kash},
can become faster than its value $c$ in vacuum,
or can even become negative~\cite{Steinberg,Wang}. 
Note that superluminal light propagation with group velocity larger than
$c$ cannot transmit
information faster than the vacuum speed of light~\cite{Chiao}, such 
that it is not at odds with causality. 
Superluminal light propagation has been investigated for many
potential uses, not only as a tool for studying a very peculiar
state of matter, but also for developing quantum computers, high
speed optical switches and communication systems \cite{Kim}.

Both experimental and theoretical studies have been performed 
to realize super- and subluminal light propagation in a
single system. For example, speed control in atomic systems has been
achieved by changing the frequencies, amplitudes or phase differences of
the applied fields. It has been shown that switching from subluminal to
superluminal pulse propagation can be achieved by the intensity of
the coupling fields~\cite{Goren,Agarwal,Han,Tajalli}, and the relative phase
between two weak probe fields~\cite{Bortman}. 
Morigi et al.~\cite{Morigi} have compared the phase-dependent
properties of the $\diamond$ (diamond) four level system with
those of the double $\Lambda$ system. 
In Ref. \cite{Wang}, gain-assisted superluminal light
propagation was observed in a cesium vapor cell while in 
most other studies, superluminal light propagation is accompanied
by considerable absorption. 
Sub- and superluminal light propagation together with nonlinear
optical gain or losses were observed in~\cite{yzhu}.
Two of the present authors suggested to
use an incoherent pump field to control light propagation from
subluminal to superluminal~\cite{Mahmoudi1, Mahmoudi2}.  
Recently, we have studied the light propagation of a probe pulse
in a four-level double lambda system, where the applied laser fields
form a closed interaction loop~\cite{Evers}. In such systems, the
finite frequency width of a probe pulse requires a time dependent
treatment of the light propagation. We have found both
sub- and superluminal light propagation without absorption or with gain,
controlled by the Rabi frequency of one of the coupling fields.

All these effects depend on the modification of the dispersive and
absorptive properties of the atomic medium. A particular class of
systems that allows to modify the optical response to a great
extend are those with so-called interacting dark 
resonances~\cite{Lukin}. A characteristic feature of such 
systems is the appearance of very sharp, high-contrast
structures in the optical spectra.
Resonances associated with double dark states can be made 
absorptive or transparent and their optical
properties such as width and position can be manipulated by
applying suitable coherent interactions.
It was also shown that very weak incoherent excitation of the atoms
can be sufficient to turn absorptive features into optical 
gain structures. This has been proposed as a model system
to obtain strong laser gain in the ultraviolet and vacuum 
ultraviolet regime by Fry et al.~\cite{Fry}.

In this paper we consider probe pulse propagation through
a system which exhibits interacting dark resonances.
The level configuration of our four-level scheme is based
on the lasing system proposed in~\cite{Fry}, and consists
of three atomic states in ladder configuration, with an 
additional fourth perturbing state coupled by a laser field to
the upper state of the ladder system. The lower transition of 
the ladder system acts as the probe transition. 
 This system can
be realized, e.g., in mercury, where the probe transition
has a low wavelength of 253.7 nm, i.e., in the ultraviolet
region.
We find that the medium susceptibility in dependence
on the probe field detuning exhibits high-contrast structures
characteristic of interacting dark states. 
These structures typically lead to 
superluminal probe field propagation with high absorption,
and thus as such are of limited  interest. If, however,
a weak incoherent pumping is applied in addition to the
probe field transition, then we find that in the region around
a narrow structure both sub- and superluminal propagation as well as
negative group velocities are possible without absorption, controlled
by the incoherent pumping strength.

\section{Analytical considerations}
\subsection{The Model System}

\begin{figure}
\centering
\includegraphics[width=15cm]{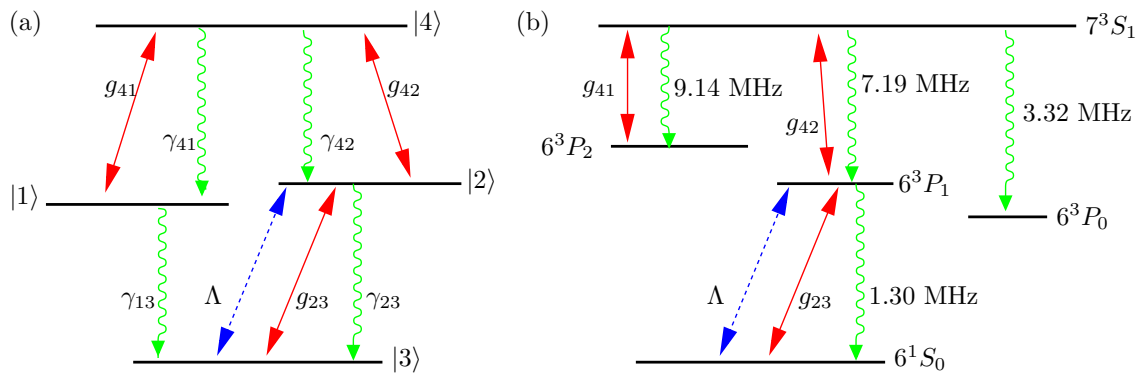}
\caption{\label{fig-scheme}(a) Energy scheme of the four level 
atomic system considered. Transition $|2\rangle \leftrightarrow |4\rangle$
is driven by a strong laser field, transition 
$|1\rangle \leftrightarrow |4\rangle$ by a weak coupling field, 
and the probe field interacts with transition 
$|2\rangle \leftrightarrow |3\rangle$. In addition, a weak 
incoherent field is applied to the probe field transition.
(b) A possible realization of the scheme
in mercury. Population transfer to
state $6{}^3P_0$ has to be compensated via a repump field.}
\end{figure}

We consider an atomic four level system as shown in figure~\ref{fig-scheme}(a).
Transition $|2\rangle \leftrightarrow |4\rangle$ is driven 
by a strong coherent field with frequency $\omega_{42}$ and
Rabi frequency $g_{42}$.
A weak coupling field with frequency $\omega_{41}$ and
Rabi frequency $g_{41}$ 
is applied to transition $|1\rangle \leftrightarrow |4\rangle$.  
The weak probe field with frequency $\omega_{23}$ and
Rabi frequency $g_{23} = g_p$
couples to transition $|2\rangle \leftrightarrow |3\rangle$.
Finally,  an incoherent driving field with pump strength
$\Lambda$ is applied to the probe transition.
We further include spontaneous decay with rates 
$\gamma_{41}$, $\gamma_{42}$, $\gamma_{23}$, and $\gamma_{13}$, respectively,
on the dipole-allowed transitions. The atomic transition
frequencies are denoted by $\bar{\omega}_{ij}$, and
the laser field detunings with respect to the atomic
transition frequencies are $\Delta_{ij} = \omega_{ij}
- \bar{\omega}_{ij}$ ($i,j\in{1,\dots,4}$).
A realization of our level scheme can be found, e.g., in mercury,
see figure~\ref{fig-scheme}(b).

The density matrix equations of motion, in the
rotating wave approximation, are
\numparts
\begin{eqnarray}
\dot{\rho}_{11} &=&
-2 \gamma_{13} \rho_{1 1} 
+ 2 \gamma_{41} \rho_{4 4}
- i g_{41}^\ast \rho_{1 4} 
+ i g_{41} \rho_{4 1} 
\,, \label{density2a}
\\ 
\dot{\rho}_{22} &=&
-2 \gamma_{23} \rho_{2 2} 
+ 2 \gamma_{42} \rho_{4 4}
- 2 \Lambda \rho_{2 2} 
+ 2 \Lambda \rho_{3 3} 
\nonumber \\
&&
+ i g_{p}^\ast \rho_{3 2} 
- i g_{p} \rho_{2 3} 
- i g_{42}^\ast \rho_{2 4} 
+ i g_{42} \rho_{4 2} 
\,,\\ 
\dot{\rho}_{33} &=&
2 \gamma_{13} \rho_{1 1} 
+ 2 \gamma_{23} \rho_{2 2} 
+ 2 \Lambda \rho_{2 2} 
- 2 \Lambda \rho_{3 3}
- i g_{p}^\ast \rho_{3 2} 
+ i g_{p} \rho_{2 3} 
\,,\\ 
\dot{\rho}_{12} &=&
-(\Gamma_{12} 
+ i \Delta_{4 1} 
- i \Delta_{4 2} + \Lambda) \rho_{1 2} 
- i g_{42}^\ast \rho_{1 4} 
- i g_{p} \rho_{1 3} 
+ i g_{41} \rho_{4 2}\,,\\ 
\dot{\rho}_{13} &=&
- ( \Gamma_{13} 
+ i \Delta_{4 1} 
- i \Delta_{4 2} 
- i \Delta_p+\Lambda)\rho_{1 3} 
-i g_{p}^\ast \rho_{1 2} 
+ i g_{41} \rho_{4 3}\,,\\ 
\dot{\rho}_{14} &=&
- (\Gamma_{14} 
+ i \Delta_{4 1} ) \rho_{1 4} 
-i g_{41} \rho_{1 1} 
+ i g_{41} \rho_{4 4}
- i g_{42} \rho_{1 2} 
\,,\\ 
\dot{\rho}_{23} &=&
- (\Gamma_{23}
- i \Delta_p 
+ 2 \Lambda) \rho_{2 3} 
-i g_{p}^\ast \rho_{2 2} 
+ i g_{p}^\ast \rho_{3 3} 
+ i g_{42} \rho_{4 3}
\,,\\ 
\dot{\rho}_{24} &=&
- (\Gamma_{24}
+ i \Delta_{4 2} 
+ \Lambda )\rho_{2 4} 
- i g_{42} \rho_{2 2} 
+ i g_{42} \rho_{4 4}
+ i g_{p}^\ast \rho_{3 4} 
-i g_{41} \rho_{2 1} 
\,,\\ 
\dot{\rho}_{34} &=&
- (\Gamma_{34} 
+ i \Delta_p 
+ i \Delta_{4 2}
+ \Lambda) \rho_{3 4} 
+ i g_{p} \rho_{2 4} 
- i g_{41} \rho_{3 1} 
- i g_{42} \rho_{3 2} 
\,,\label{density2j}
\\ 
\rho_{44}&=& 1-\rho_{11}-\rho_{22}-\rho_{33}\,.
\end{eqnarray}
\endnumparts
In the above equations, $\Gamma_{ij}=(2\gamma_i+2\gamma_j)/2$ are the 
damping rates of the coherences with $\gamma_i$ being the total
decay rate out of state $|i\rangle$, and $\Delta_p = \Delta_{23}$ is the
probe field detuning.

Our main observable is the response of the atomic medium to the
probe field. As will be discussed in Sec.~\ref{sec-obs}, the linear 
susceptibility of the weak probe field 
is determined by the probe transition coherence
$\rho_{23}$.
We therefore proceed by solving the above 
equations~(\ref{density2a})-(\ref{density2j})
in the steady state under the assumption of specific
parameter relations. 

First, in the absence of the incoherent pump field ($\Lambda = 0$), 
an expansion of the steady state coherence $\rho_{23}$ to the leading
order in the probe field Rabi frequency $g_p$ yields
\numparts
\begin{eqnarray}
\rho_{23}&=&\frac{- g_p (|g_{41}|^2-C_{13} \cdot C_{34})}
{ |g_{41}|^2 C_{23}+ C_{13} ( |g_{42}|^2 - C_{23} \cdot C_{34})}\,, \label{analytic1}\\
C_{13} &=& \Delta_p - \Delta_{41} + \Delta_{42} + i \,\Gamma_{13} \,,\\
C_{34} &=&  \Delta_p + \Delta_{42} + i \,\Gamma_{34}\,,\\
C_{23} &=& \Delta_p + i \,\Gamma_{23}\,.
\end{eqnarray}
\endnumparts
It will turn out that an interesting parameter range for the present study
is given by
\numparts
\begin{eqnarray}
\Delta_{41} &=& \Delta_{42} = 0 \,, \label{condition1a}\\
\Delta_{p} &\ll& \gamma_{31}, \gamma_{41}, \gamma_{42} \,,\\
g_{41} &\ll& g_{42} \,,\\
\gamma_{13} &=& 0\,. \label{condition1d}
\end{eqnarray}
\endnumparts
In this limit, equation~(\ref{analytic1}) becomes
\begin{equation}
\label{approx}
\rho_{23}=\frac{- g_p (|g_{41}|^2- i \Delta_{p}\Gamma_{34})}
{|g_{42}|^2 \Delta_p +i [|g_{41}|^2\Gamma_{23}-\Delta_p^2(\Gamma_{34}+\Gamma_{23})]}\,.
\end{equation}
An inspection of equation~(\ref{approx}) reveals that the imaginary part is
strictly positive, and the half width of the absorption peak around 
$\Delta_p=0$ is
determined by
\begin{equation}\label{width}
w\simeq \left(\frac{g_{41}}{g_{42}}\right)^2\Gamma_{23} 
= \left(\frac{g_{41}}{g_{42}}\right)^2\gamma_{23}\,.
\end{equation}

Next, we seek the corresponding steady state solution for $\rho_{23}$ with incoherent
pump field with pump intensity $\Lambda$. The parameters are chosen to satisfy
equations~(\ref{condition1a})-(\ref{condition1d}) as 
well as the new condition on the pump field
\begin{equation}
\label{condition2}
\Lambda_0  \ll \Lambda  \ll \gamma_{41},\gamma_{42} \,.
\end{equation}
Further, we assume the Rabi frequencies $g_{ij}$ to be real in the
following. We obtain in leading order of the probe field
coupling $g_{p}$
\begin{equation}\label{analytic2}
\rho_{23}=\frac{g_{41}^2\, g_{p}\,\gamma_{23}}{(g_{42}^2\,
\gamma_{23}+2\Lambda\,\Gamma_{24}\,\gamma_{42})}
\frac{\Delta_{p}-i\Lambda}{\Delta_{p}^2+\Lambda^2}\,.
\end{equation}
Here the parameter $\Lambda_0$ is defined by 
\begin{equation}
\Lambda_0= \frac{g_{41}\, \gamma_{23} \,(\gamma_{41}+\gamma_{23})}{g_{42}^2
\,\gamma_{41}+\gamma_{23}\, \Gamma_{34} \,(\gamma_{41}+\gamma_{23})}\simeq
\left ( \frac{g_{41}}{g_{42}}\right )^2\gamma_{23}\,.
\end{equation}
Since $|g_{41}/g_{42}|^2\gamma_{23}$ can be made small, 
for a suitable combination of the Rabi frequencies $g_{41}$ and
$g_{42}$ the condition $\Lambda \gg \Lambda_0$ can be fulfilled
even for incoherent pump strengths which are orders of magnitude  
smaller those required, e.g., to saturate the optical 
transition.

We find that the imaginary part of equation~(\ref{analytic2})
is negative if the condition $\Lambda \gg \Lambda_0$
is fulfilled.  Thus $\Lambda_0$ indicates the incoherent 
pumping rate at which the absorption peak turns into a gain
structure, if the conditions in
equations~(\ref{condition1a})-(\ref{condition1d}) and (\ref{condition2})
are fulfilled.

\subsection{\label{sec-obs}Observables}

Our main observable is the response of the atomic medium to the
probe field. The linear susceptibility of the weak probe field can 
be written as~\cite{scullybook}
\begin{equation}\label{sus}
\chi (\omega_p)=\frac{2N \eta_{p}}{\epsilon_0
E_{p}}\rho_{23}(\omega_p) \,,
\end{equation}
where $N$ is the atom number density in the medium, $\eta_p$ is
the probe transition dipole moment and
$\chi=\chi^\prime+i\chi^{\prime\prime}$. The real and imaginary
parts of $\chi(\omega_p)$ correspond to the dispersion and the
absorption, respectively.  
The slope of the dispersion with respect to the probe detuning 
has a major role in the calculation
of the group velocity. We introduce the
group index, $n_g=c/v_g$, where the group velocity $v_g$ of the
probe field is given by~\cite{Hau,Wang}
\begin{equation}\label{group}
v_g=\frac{c}{\left [1+2\pi\chi^\prime
(\omega_p)+2\pi \omega_p \frac{\partial\chi^\prime(\omega_p)}{\partial
\omega_p}\right]}\,.
\end{equation}
Equation~(\ref{group}) implies that, for a negligible real part $\chi^\prime(\omega_p)$, 
the group velocity can be significantly reduced via a steep positive
dispersion. Strong negative dispersion, on the other hand,  
can lead to an increase in 
the group velocity and even to a negative group velocity.

Substituting equations~(\ref{analytic2}) and (\ref{sus}) in
equation~(\ref{group}), the group index of the probe field 
evaluates to
\begin{equation}\label{index}
n_g-1=\frac{g_{41}^2 \, g_{p}\, \gamma_{23}}
{(g_{42}^2 \, \gamma_{23}+2\,\Lambda\,\Gamma_{24}\,\gamma_{42})}
\frac{\Delta_{p}^2-\Lambda^2}{(\Delta_{p}^2+\Lambda^2)^2}\,.
\end{equation}
It can be expected from equation~(\ref{index}) that for suitable parameters,
the group index around $\Delta_p=0$ is negative and accompanied by 
gain, and this is indeed what we find below. 

The relation between coherence and susceptibility equation~(\ref{sus})
can be rewritten as
\begin{eqnarray}
\chi (\omega_p) &=&\frac{2N \eta_{p}}{\epsilon_0 E_{p}}\rho_{23}(\omega_p) 
= \frac{3 N  \lambda_{p}^3}{4\pi^2} \:\frac{\gamma_{23}}{\gamma} \: 
\frac{\rho_{23}(\omega_p)}{g_{p}/\gamma} \,,
\end{eqnarray}
where we have used $\gamma_{23} = (\eta_{p}^2
\bar{\omega}_{23}^3)/(3\pi\epsilon_0 \hbar c^3)$ and 
$g_{p} = \eta_{23}E_{p}/\hbar$ as well as $\omega_{23} = 2\pi 
c/\lambda_{23}$ with the probe transition wavelength $\lambda_{23}$.
For mercury probe wavelength 253.7 nm, particle density $N=10^{12}$cm$^{-3}$
and $\gamma_{23}/\gamma = 0.14$ as found in mercury one finally obtains
\begin{equation}
\chi (\omega_p) = 1.74\times 10^{-4}\:
\frac{\rho_{23}(\omega_p)}{g_{p}/\gamma}
 \,.
\end{equation}
Throughout our discussion of numerical results,
we will assume these parameters in order to evaluate the susceptibility.

\subsection{\label{sec-dressed}Dressed-state analysis}

We now introduce the dressed states generated by the  strong driving
field acting on transition $|2\rangle \leftrightarrow |4\rangle$ 
and the coupling field acting on transition
$|1\rangle \leftrightarrow |4\rangle$, in order to demonstrate
the presence of interacting dark resonances due to the
perturbing field with Rabi frequency $g_{41}$~\cite{Fry}.
In the absence of the incoherent pump field, the dressed states are
\numparts
\begin{eqnarray}
|0\rangle&=&-\frac{g_{42}}{\sqrt{g_{41}^2+g_{42}^2}}\,|1\rangle+
\frac{g_{41}}{\sqrt{g_{41}^2+g_{42}^2}}\,|2\rangle \,, \\
|\pm\rangle&=&\frac{g_{41}}{\sqrt{2(g_{41}^2+g_{42}^2)}}\,|1\rangle+
\frac{g_{42}}{\sqrt{2(g_{41}^2+g_{42}^2)}}\,|2\rangle 
\mp\frac{1}{\sqrt{2}}\,|4\rangle \,,
\end{eqnarray}
\endnumparts
with energies
\begin{eqnarray}
\lambda_0&=&0 \,, \qquad
\lambda_\pm =\pm \hbar \sqrt{g_{41}^2+g_{42}^2}.
\end{eqnarray}
The two dressed states $|\pm\rangle$ correspond, in the limit of
vanishing driving field $g_{41}$, to the
usual Autler-Townes dressed components split by $2\hbar g_{42}$.
The third dressed state $|0\rangle$ coincides in this limit with
the bare state $|1\rangle$ and hence is decoupled  from the
fields. This is no longer so in the presence of a second weak
driving field $g_{41}$. In this case the dressed state
$|0\rangle$ contains an admixture of $|2\rangle$ and thus has a
nonzero dipole matrix element with the state $|3\rangle$. As a
result of this coupling, there are transitions between $|0\rangle$
and $|3\rangle$, corresponding to three photon resonances from
$|1\rangle$ to $|3\rangle$ that exhibit interference 
effects~\cite{Fry}.

%
%
\begin{figure}[t]
\centering
\includegraphics[width=8cm]{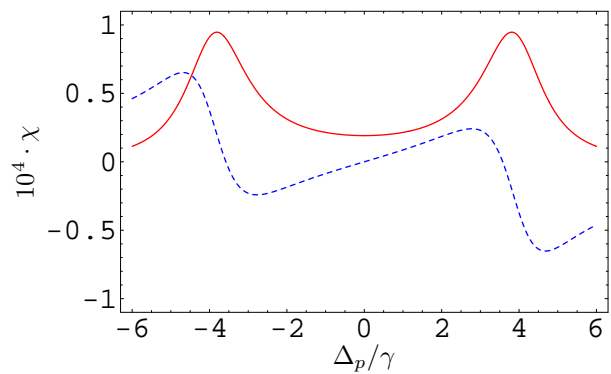}
\caption{\label{fig-3}Real (blue dashed) and imaginary (red solid)
parts of the susceptibility $\chi$ as a function of the probe 
detuning $\Delta_p$ for the parameters 
$\gamma_{41}=\gamma$, $\gamma_{23}=0.14\,\gamma$,
$\gamma_{42}=0.79\,\gamma$, $\gamma_{13}=0.01\,\gamma$,
$g_{p}=10^{-4}\,\gamma$, $g_{41}=0$, $g_{42}=4\,\gamma$, 
$\Lambda=0$, $\Delta_{42}=\Delta_{41}=0$.}
\end{figure}

\section{\label{sec:res}Results}

In figure~\ref{fig-3} we show the real (blue dashed) and 
imaginary (red solid) part of the probe field susceptibility
$\chi$ versus the probe detuning $\Delta_p$, 
which correspond to the dispersive and absorptive properties
of the medium, respectively. 
In this figure, the perturbing laser field is switched off, $g_{41}=0$.
The other parameters are 
$\gamma_{41}=\gamma, \gamma_{23}=0.14\,\gamma$,
$\gamma_{42}=0.79\,\gamma$, $\gamma_{13}=0.01\,\gamma$,
$g_{p}=10^{-4}\,\gamma, g_{42}=4\,\gamma$, $\Lambda=0$, $\Delta_{42}=\Delta_{41}=0$.
Note that the ratios of the decay rates correspond to the case 
found in mercury, see figure~\ref{fig-scheme}(b).
We have added a weak decay rate $\gamma_{13}$, since otherwise
in the steady state all population is trapped in $|3\rangle$.
The driving field with Rabi frequency $g_{42}$ leads to an Autler-Townes 
doublet with a dip in the absorption at zero
detuning, i.e., partial electromagnetically induced transparency (EIT). 
The slope of the real part of the susceptibility in the region
of reduced absorption is positive. We thus find that subluminal 
light propagation occurs around 
zero detuning with reduced absorption as it is common for EIT. 
If the state $|4\rangle$ was long-lived, then the EIT  
leading to the partial transparency would be more pronounced such that
the absorption would vanish at zero detuning.

\begin{figure}[t]
\centering
\includegraphics[width=7.5cm]{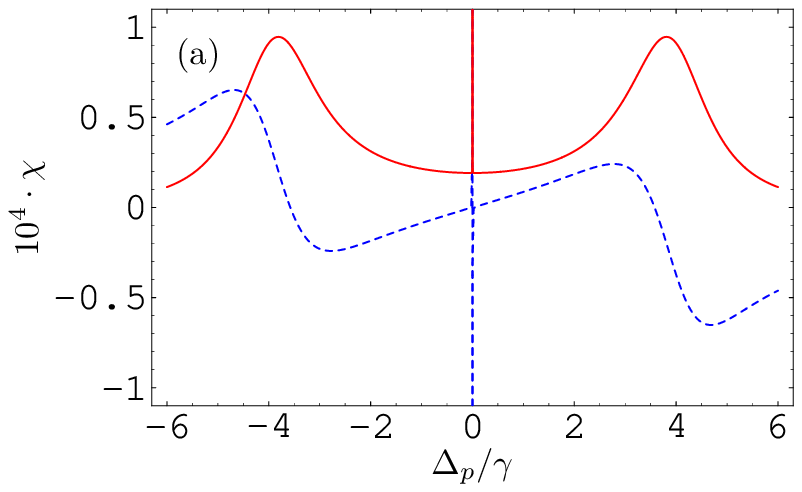}
\hspace*{0.2cm}
\includegraphics[width=7cm]{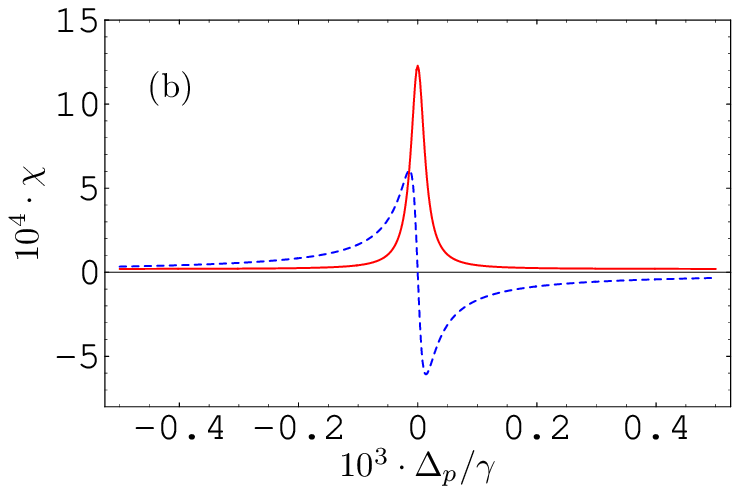}
\caption{\label{fig-4}Real (blue dashed) and imaginary (red solid)
parts of the susceptibility $\chi$ as a function of the probe detuning
for $\gamma_{13}=0$ and $g_{41}=0.04\, \gamma$.
The other parameters are the same as in figure~\ref{fig-3}. 
(b) is a closeup on the central part of (a).}
\end{figure}

In figure~\ref{fig-4}, in addition we apply the weak perturbing 
field with Rabi frequency $g_{41}=0.04\,\gamma$, and assume negligible
decay on transition $|3\rangle \leftrightarrow |1\rangle$, since a trapping
in this state is now avoided by the additional laser field.
The results are identical to figure~\ref{fig-3} except for a narrow 
absorption spike at around zero detuning. The shape and width of the absorption 
spike are determined by equation~(\ref{analytic1}) and equation~(\ref{width}), 
respectively. In particular, the width is much less than the natural linewidth.
Again, for a long-lived state $|4\rangle$, the transparency regions on each
side of the absorption spike would become two points of EIT, i.e.,
a double dark state~\cite{Lukin}. 
In terms of the light propagation, the
slope of the real part of the susceptibility around zero detuning is negative 
such that  superluminal light propagation could be observed, albeit with high 
absorption.

\begin{figure}[t]
\centering
\includegraphics[width=7.5cm]{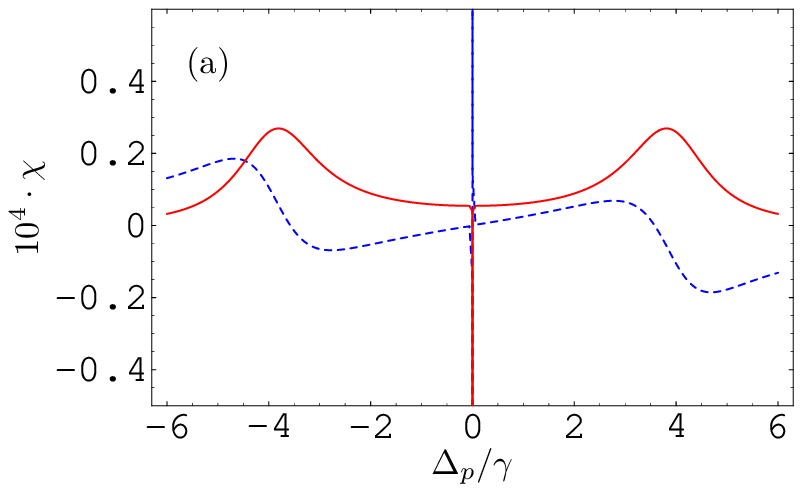}
\hspace*{0.2cm}
\includegraphics[width=7.5cm]{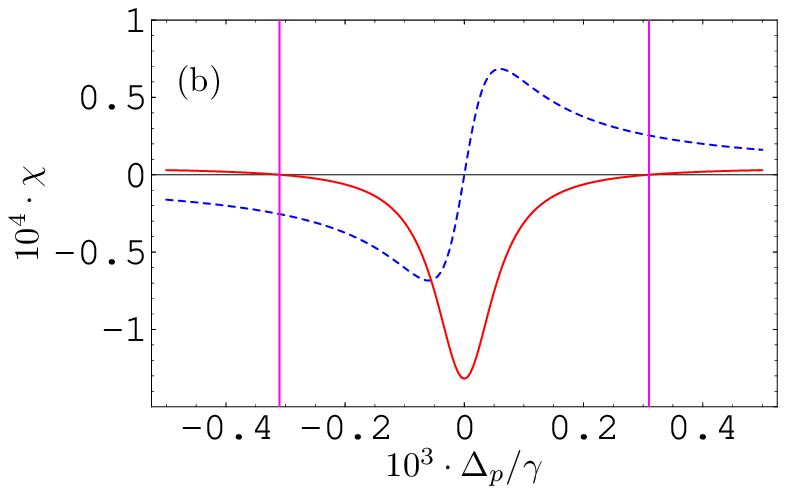}
\caption{\label{fig-5}Real (blue dashed) and imaginary (red solid)
parts of the susceptibility $\chi$ as a function of the probe detuning
for $\Lambda=4\times 10^{-5}\,\gamma$. The other parameters 
are the same as in figure~\ref{fig-4}.
(b) is a closeup on the central part of (a). The purple vertical lines indicate the
roots of the imaginary part.}
\end{figure}

We now in addition apply a weak incoherent pumping field on the probe 
transition $|2\rangle-|3\rangle$. 
Figure~\ref{fig-5} shows the corresponding results.
The incoherent pump field rate is chosen as $\Lambda=4\times 10^{-5}\gamma$.
In this case, the superluminal light propagation found in figure~\ref{fig-4}
at $\Delta_p=0$ switches to subluminal propagation, and the absorption spike
at zero detuning becomes a gain spike. The shape of the gain
spike is determined by the imaginary part of equation~(\ref{analytic2})
which is Lorentzian with halfwidth equal to $\Lambda$. 
This can be understood from the fact that the
spike arises from a three-photon transition from
$|1\rangle$ to $|3\rangle$. In the absence of the incoherent pump
field, most of the population is in level $|3\rangle$. Therefore,
the probe field is absorbed. But in the presence of the weak incoherent
pump field, the population is transferred to level $|1\rangle$.
This optical pumping leads to the gain spike in the spectrum.

Furthermore, it can be seen from Fig.~\ref{fig-5} that at 
$\Delta_p \approx \pm 3.1 \times 10^{-4} \gamma$ (indicated by the
purple vertical lines), the imaginary
part of the susceptibility vanishes together with a negative slope
of the real part. At these probe field detunings, the
real part of the susceptibility itself is non-zero, and is negative
(positive) for $\Delta_p \approx - 3.1 \times 10^{-4} \gamma$
($\Delta_p \approx - 3.1 \times 10^{-4} \gamma$).
In the following, we discuss the two cases of interest
with resonant or non-resonant probe field separately.

\begin{figure}[t]
\centering
\includegraphics[height=4.5cm]{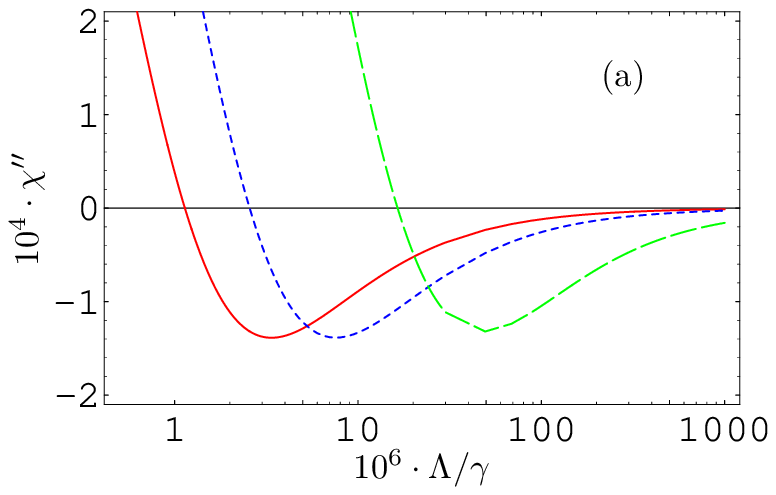}
\hspace*{0.5cm}
\includegraphics[height=4.5cm]{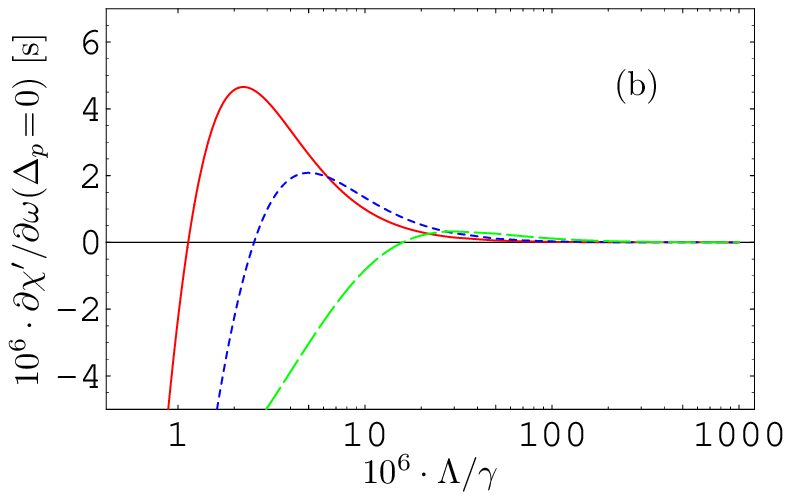}
\caption{\label{fig-lambda}(a) Imaginary part $\chi^{\prime\prime}$
of the susceptibility as a function of the pumping field
strength $\Lambda$. The parameters are as in figure~\ref{fig-4} with
$\Delta_p = 0$, and $g_{42}=15\gamma$ (solid red), $10\gamma$ 
(short-dashed blue),
$4\gamma$ (long-dashed green). (b) Slope of the real part $\chi^{\prime}$
of the susceptibility at zero probe field detuning $\Delta_p = 0$
for parameters as in (a).}
\end{figure}

We start with the resonant case $\Delta_p = 0$.
In figure~\ref{fig-lambda}(a), we study the effect of the incoherent
pumping strength $\Lambda$ on the magnitude of the imaginary
part of the susceptibility $\chi$ at resonance
$\Delta_p = 0$. It can be seen that depending on the
coupling field Rabi frequency $g_{42}$, the transition 
from absorption to gain occurs at 
different values of the incoherent pumping. For the parameters
of figure~\ref{fig-4}, which correspond to the long-dashed green curve in 
figure~\ref{fig-lambda}(a), the transition is at about
$\Lambda\approx 2\times 10^{-5}\,\gamma$. This explains
why gain could be observed for the parameters in
figure~\ref{fig-5}. On increasing 
the incoherent pumping further, the imaginary part approaches zero again.

%

After the discussion of the absorption, we now turn to a discussion
of our main observable, the group velocity.
Since the real part of $\chi$ itself vanishes at
$\Delta_p=0$, the group velocity is determined by the
slope of the real part of the susceptibility at
$\Delta_p = 0$, see equation~(\ref{group}). This quantity is shown in 
figure~\ref{fig-lambda}(b).
It can be seen that, for no or small incoherent pumping, the system
exhibits a negative slope, which leads to a superluminal or even negative 
group velocity.
On increasing $\Lambda$, the 
slope can be adjusted to large positive values, where subluminal
light can be expected. Thus in principle the
system allows for a wide range of group velocities, controlled
via the incoherent pump rate $\Lambda$. 
But from a comparison of figures~\ref{fig-lambda} (a) and (b) it
can be seen that typically negative slopes are accompanied by
absorption, while positive slopes occur together with gain.
Thus at $\Delta_p \approx 0$, only a reduction of the group
velocity is accessible in experiments without absorption.
The different curves in figure~\ref{fig-lambda} further show 
that the precise response of
the system to the incoherent pumping can be controlled by varying
the coupling field Rabi frequency $g_{42}$. In particular, stronger
coupling fields $g_{42}$ may be favourable, since then
the range of possible slopes is increased, as can be seen from 
figure~\ref{fig-lambda}(b).

\begin{figure}[t]
\centering
\includegraphics[height=4.5cm]{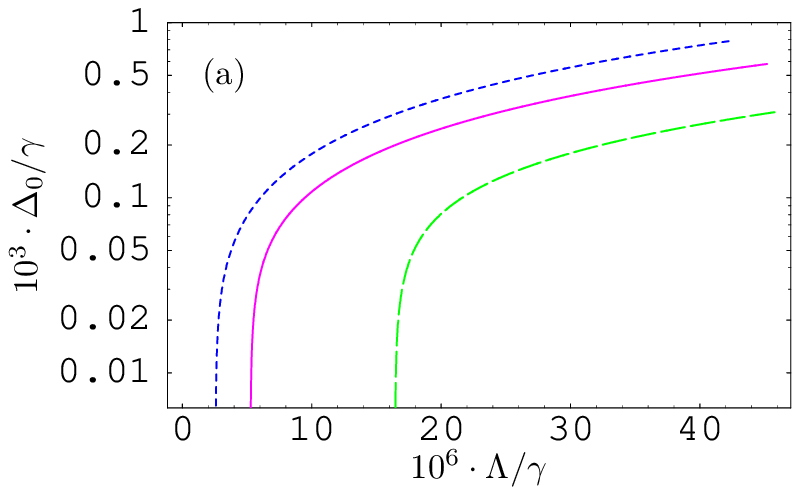}
\hspace*{0.5cm}
\includegraphics[height=4.4cm]{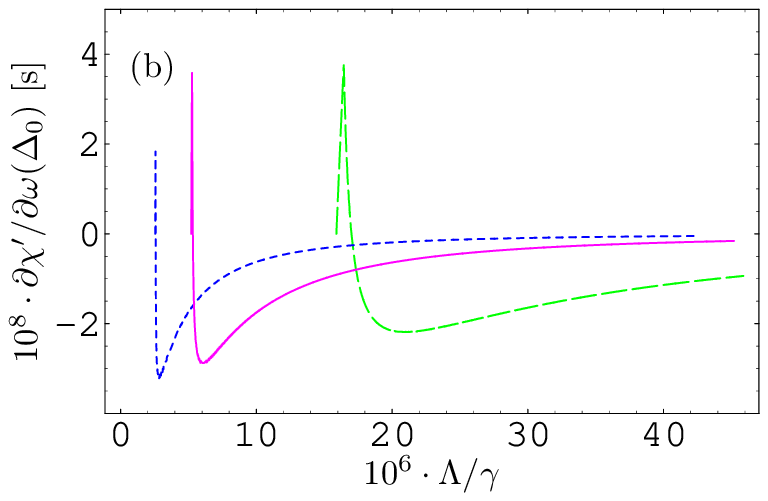}
\caption{\label{fig-offres}(a) Position $\Delta_0$ of the root of the imaginary
part of the susceptibility $\chi$ as a function of the pumping field
strength $\Lambda$. At this frequency, light passes unattenuated 
through the medium. (b) Slope of the real part $\chi^{\prime}$
of the susceptibility at the detuning $\Delta_0$ with vanishing absorption.
The parameters are as in figure~\ref{fig-4} but with
$g_{42}=10\gamma$  (short-dashed blue), $g_{42}=7\gamma$ (solid purple), and
$g_{42}=4\gamma$ (long-dashed green). }
\end{figure}

We now turn to a discussion of the non-resonant case, $\Delta_p \neq 0$,
and focus on the regions with vanishing absorption, such as 
$\Delta_p \approx \pm 3.1 \times 10^{-4} \gamma$
in figure~\ref{fig-5}. It can be seen that around these probe field
detunings, the imaginary part of the susceptibility vanishes, such that
the probe field passes unattenuated through the medium. At the same
time, the real part of the susceptibility is non-zero, and has a negative
slope. Therefore, at these frequencies, superluminal or negative
group velocities are accessible without absorption.
In order to study this result in more detail, in figure~\ref{fig-offres}(a)
we show the probe field detuning $\Delta_0$ at which the imaginary part 
of the susceptibility vanishes
as a function of the incoherent pumping rate $\Lambda$.
It can be seen that for no or small incoherent pumping $\Lambda$, there is always absorption
such that no $\Delta_0$ can be found. Once $\Lambda$ is large enough for a
root in the imaginary part of the susceptibility to occur, the position of the
root first increases rapidly with $\Lambda$, and then saturates. The required
value of $\Lambda$ also depends on the strength of the coupling field
$g_{42}$ as can be seen from figure~\ref{fig-offres}(a).

The corresponding figure~\ref{fig-offres}(b) depicts the slope of the real part of the susceptibility
as a function of $\Lambda$. It can be seen that by varying the pump field strength $\Lambda$,
both positive and negative slopes can be achieved at frequencies where the medium absorption
is zero. After passing through a maximum positive slope, the slope drops to a minimum
negative slope and then slowly increases again towards vanishing slope.
For every value of the coupling field Rabi frequency $g_{42}$, optimum values of 
$\Lambda$ can be identified where the slope is steepest and either positive or negative.
The maximum absolute values of the slope are of order $10^{-8}$ s$^{-1}$, such that
the third term $2\pi \omega_p \partial \chi^{\prime}/\partial \omega$ in
the denominator of equation~(\ref{group}) for our probe transition
varies between approximately $-10^9$ and $+10^9$. Therefore, strongly sub- and superluminal
propagation as well as a large range of negative group velocities occur without
absorption in our sample, controlled by the magnitude of the incoherent pumping.
It should be noted that only very weak incoherent pumping is required, as can
be seen from the scaling of the x-axes in figures~\ref{fig-offres}.

\begin{figure}[t]
\centering
\includegraphics[width=12cm]{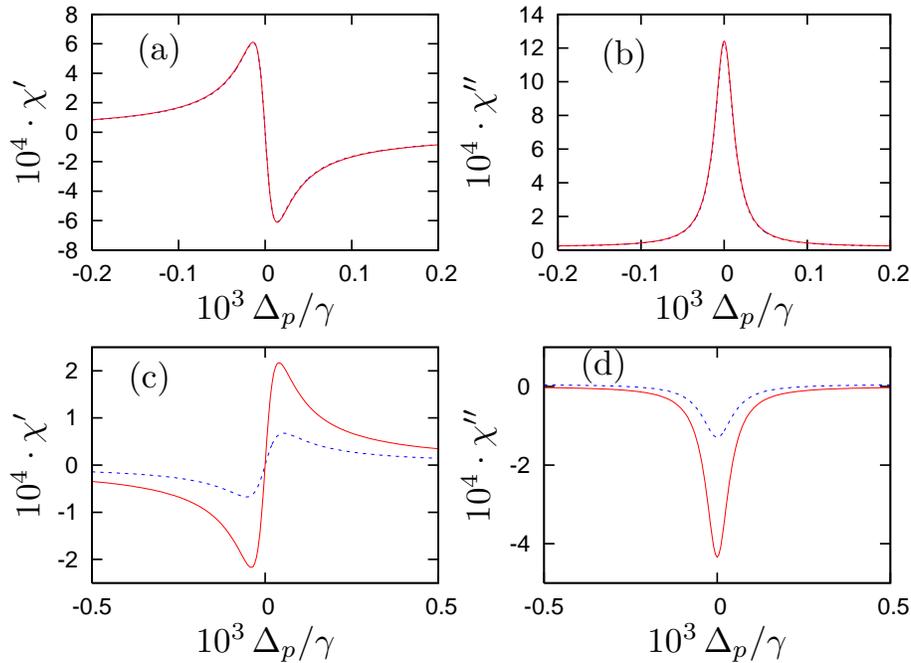}
\caption{\label{fig-6}Real (a,c) and imaginary (b,d)
parts of the susceptibility $\chi$ as a function of the probe detuning.
The analytical results are shown as solid red lines,
whereas our numerical results are shown as dashed blue
lines. The parameters in (a,b) are as in figure~\ref{fig-4},
and equation~(\ref{approx}) is shown as the analytical result.
The parameters in (c,d) are as in figure~\ref{fig-5},
with equation~(\ref{analytic2}) as analytical result.}
\end{figure} 

Throughout this section, the figures~\ref{fig-3}-\ref{fig-offres}
have been obtained from a numerical solution of the full
density matrix equations~(\ref{density2a})-(\ref{density2j}). 
In the following, we 
verify our approximate  analytical expressions,
equations~(\ref{analytic1})-(\ref{analytic2}), by a comparison to the exact
numerical calculations. The result is shown in figure~\ref{fig-6},
where the solid red curves correspond to the approximate analytical solutions,
whereas the blue dashed curves represent our numerical
results.
The approximate result
equation~(\ref{approx}) for the case without incoherent pump field is
shown in comparison to the numerical data in figure~\ref{fig-6}(a,b).
It turns out that in this case, the results from 
equation~(\ref{approx}) are virtually identical
to the corresponding numerical results. 
Equation~(\ref{analytic2}) for the case with incoherent pumping is
compared to the numerical results in figure~\ref{fig-6}(c,d).
Here, the analytic results only describe the
qualitative behavior of the curves. The reason for this is that in this
figure, we chose parameters for which the condition $\Lambda \gg \Lambda_0$ 
in equation~(\ref{condition2}) is not well satisfied. If the incoherent
pumping $\Lambda$ is increased, the agreement of the approximate
results with the numerical calculation improves.
Thus we conclude that our analytical results describe the
system well enough to allow for an optimization of
the parameters towards a desired peak structure, as long
as the conditions on the parameters are satisfied.

\section{\label{sec:conc}Conclusion}
We have discussed the dispersive and absorptive properties 
of a four-level atomic medium that exhibits interacting
dark-state resonances. In our numerical analysis, we have
focused on mercury atoms with an ultraviolet  probe field wavelength
of 243.7 nm. Due to the interacting resonances,
a high-resolution structure appears in both the absorption
and the dispersion spectra. A weak probe field 
tuned to this resonance usually experiences superluminal
propagation with absorption. But if in addition a 
weak incoherent pump field is applied to the probe transition, 
then the superluminal light propagation changes to  
subluminal light propagation accompanied by no absorption or gain.
Slightly off resonance, the probe field experiences a
vanishing imaginary part of the susceptibility.
At these off-resonant frequencies, the real part of the susceptibility 
itself is non-zero and has a slope depending on the incoherent 
pumping strength. Thus both sub- and superluminal light
propagation as well as negative group velocities can be
achieved without absorption.
The control via the incoherent pump fields suggests
potential applications, e.g., in optical switching devices or 
in controllable pulse delay lines for the ultraviolet frequency
region.

\ack
MM gratefully acknowledges support for this work from the German
Science Foundation and from Zanjan University, Zanjan, Iran.

\end{document}